\documentclass[manuscript,trackchanges]{aastex62}

\begin{document}

\title{N-Body Simulations of Gas-free Disc Galaxies with SMBH Seed in Binary Systems}

\author{R. Chan}

\affil{Observat\'orio Nacional,
Coordena\c c\~ao de Astronomia e Astrof\'{\i}sica,
Rua General Jos\'e Cristino 77, S\~ao
Crist\'ov\~ao, CEP 20921--400, Rio de Janeiro, RJ, Brazil. 
}
\email{chan@on.br}

\begin{abstract}
We have shown the outcome of N-body simulations of the 
interactions of two disc galaxies without gas with the same mass. Both disc galaxies have 
halos of dark matter, central bulges and initial supermassive black hole (SMBH) 
seeds at their centers. 
The purpose of this work is to study the mass and dynamical evolution 
of the initial SMBH seed during a Hubble cosmological time. It is a complementation of our previous paper
with different initial orbit conditions and by introducing the SMBH seed in the initial galaxy.
The disc of the secondary galaxy has coplanar or polar orientation in relation 
to the disc of the primary galaxy and their initial orbit are eccentric and prograde.  
The primary and secondary galaxies have mass and size of Milky Way with an initial SMBH seed. 
We have found that the 
merger of the primary and secondary discs can result in a final normal disc or a final warped disc.
After the fusion of discs,
the final one is thicker and larger than the initial disc.
The tidal effects are very important, modifying the evolution of the SMBH 
in the primary and secondary galaxy differently.
The mass of the SMBH of the primary galaxy have increased by a factor ranging 
from 52 to 64 times 
the initial seed mass, depending on the experiment. However,
the mass of the SMBH of the secondary galaxy have increased by a factor ranging 
from 6 to 33 times the initial SMBH seed mass, depending also on the experiment. 
Most of the accreted particles have come from the bulge and from the halo,
depleting their particles. This could explain why the observations show
that the SMBH with masses of approximately $10^6 M_\odot$ are found in
many bulgeless galaxies.
Only a small number of the accreted particles has come from the disc.
In some cases of final merging stage of the two galaxies, 
the final SMBH of the secondary galaxy was {ejected out of the galaxy}. 
\end{abstract}


\section{Introduction}

It is well known in the literature that supermassive black holes (SMBH) exist in the majority
of the galaxies, within elliptical, disc to even in dwarf galaxies 
{[}\cite{Kormendy2013}, \cite{Moran2014}{]}.

Several recent works in numerical simulations with SMBH with gas
{[}\cite{Tremmel2018}, \cite{Sanchez2018}, \cite{Curd2018}{]} {show us} {how complex is} the dynamical evolution
and mass growing with gas accretion can be.

{
Moreover, many papers have been published about simulations of binary mergers with BH seeds including complex dissipative processes but not included in the present simulations
[\cite{Springel2005}, \cite{Matteo2008}, \cite{Khan2018}, \cite{Gabor2016},\cite{Callegari2009},
\cite{Hopkins2005}, \cite{Hopkins2006}, \cite{Chapon2013}, \cite{Mayer2007}].
}

{
Simulations of binary mergers with BH seeds and no dissipative effects, similar to the ones presented in this work are published by several authors [\cite{Li2017}, \cite{Governato1994},\cite{Ebisuzaki1991}, \cite{Makino1993}, \cite{Makino1996},  \cite{Khan2012}, 
\cite{Rantala2018}, \cite{Makino1997}].
}

On the other hand, there are only few works in the literature based on simulations of
interaction of gas-free disc galaxies 
{[}\cite{Oh2008}, \cite{Dobbs2010},\cite{Lotz2010}, \cite{Struck2011} \cite{Bois2011}, \cite{Chan2003}, \cite{Chan2014}{]}
but none has treated the problem of the existence of a SMBH at center of the galaxies.

{In a recent} paper of Li et al. (2017) 
{it is} 
presented the results of the {gas-free} interaction of SMBHs 
in very eccentric galaxy orbits. 
Besides, there are rare works studying the evolution of such binary galaxy in a 
long interval of time {[} \cite{Chan2001}, \cite{Chan2003}, \cite{Chan2014}{]}, in small eccentricity orbits.
This work is a complementation of our previous work \cite{Chan2014}, where, there, the focus was the evolution of the discs,
but here we use different initial orbit conditions and
with a SMBH seed in the initial galaxy. Thus, differently, we will focus in the SMBH seed evolution
in a cosmological time {and covering a wider range of orbits
of the galaxy binary than in the work of 
Li et al. (2017).}

Thus, the main goal of the present work is to perform numerical N-body 
simulations to study the time evolution of the mass and dynamics of the initial SMBH seed
in the two disc galaxies. We also want to know if the tidal forces
affect the evolution of the SMBH.

This paper explores the scenario 
as follows:  first, we assume a disc galaxy with the characteristics of {the
Milky Way} (disc, bulge, halo and SMBH).  Second, we let a secondary galaxy 
with the same characteristics orbit on prograde coplanar or polar
disc (orientation in relation to the primary disc galaxy). 

The paper is organized as follows: in Section 2 we describe the numerical 
method used in the simulations.
In Section 3 we present the initial conditions.  
In Section 4 we describe the results of the simulations.
Finally, in Section 5 we summarize the results.

\section {Numerical method}

The N-body simulation code used was GADGET \cite{Springel2001}.
A modified version of this basic code was made in order to introduce the SMBH 
gravitational interaction with the other particle.
Here we have used only the N-body integration but without gas.

The units used in all the simulations are $G=1$, [length] = $4.500$ kpc, [mass] =
$5.100\times 10^{10} M_\odot$, [time] = $1.993\times 10^{7}$ years
($H_0 = 100$ km/s/Mpc) and [velocity] = 220.730 km/s.
Hereinafter, all the physical quantities will be referred to these units.
The Hubble time $t_H$ corresponds to 490 time units.
We assumed in all the simulations the tolerance parameter $\theta=0.577$.
The energy is
conserved to better than 6\% during the entire evolution with a time step size
$\Delta t = 1.000\times 10^{-3}$ and the softening parameter $\epsilon=8.000\times10^{-4}$.

As mentioned above, we have utilized in this work a modified version of the 
GADGET code \cite{Springel2001},
in order to mimic the interaction of the galaxy particles with the SMBH particle.
We have assumed that the collisions between the galaxy particles and the SMBH particle
are inelastic. The collision is in such way that they fuse {with} the SMBH 
particle with the total mass as a sum of the two ones. 

The Schwarzschild radius of the SMBH is define as
\begin{equation}
R_{bh}=\frac{2 M_{bh}}{c^2},
\label{Rbh}
\end{equation}
where  $M_{bh}$ is the SMBH mass and $c$ is the light velocity. 

We also have assumed if a galaxy particle with softening parameter $\epsilon$
and it grazes the Schwarzschild radius of the SMBH, following the Equation (\ref{D}), 
they are merged {with} one single
SMBH particle (see Figure \ref{Rbh}). Thus, the Schwarzschild radius increases because of 
the additional merged galaxy mass particle {with} the SMBH.

Definition of the condition when there is a merge between 
the galaxy and the SMBH particle
\begin{equation}
D \le R_{bh} + \epsilon,
\label{D}
\end{equation}
where $D$ is the distance between the centers of the SMBH and the galaxy particle,
$\epsilon$ is
the particle softening parameter and $R_{BH}$ is the Schwarzschild radius of the SMBH.

This is clearly an oversimplified scenario of accretion of galaxy mass {onto} a SMBH
which has, in reality, a much more complex physics involved. At least, with this study, 
we can know approximately the evolution of the SMBH mass and its dynamical
evolution in binary galaxies during a long-time evolution.

{
In order to determine the number of bulge/halo/disk particles accreted onto
the SMBH as a function of time we have saved a snapshot at each time step 
of 4.9 time units until the Hubble time $t_H=490$ of both galaxies, 
primary and secondary.
Thus, at the end of each experiment we have 100 saved snapshot files.
Each snapshot file generated by the modified GADGET code has an
identification number, position, velocity and mass of each particle.
In this way we can identify the bulge/halo/disk galaxy structure that each
particle belongs. When a given particle is merged with the SMBH, using
the condition (\ref{D}), we sum the mass of this particle
with the previous mass of the SMBH and set zero mass to this particle. Besides,
we recalculate the new position and velocity of the SMBH after the
inelastic collision and then we let evolve the system again.
At the end of each experiment we count how many bulge/halo/disk
particles with zero mass that certainly have merged with the SMBH. 
Thus, we can obtained
the time evolution of the bulge/halo/disk accreted {onto} the SMBH.
}

\section{Initial conditions of the simulations}

We have utilized the self-consistent disc-bulge-halo
galaxy model by Kuijken \& Dubinski \cite{Kuijken1995} in the 
simulations, the same as in our previous paper 
(see Table \ref{table1} and Table \ref{table1a})\cite{Chan2014}. 
We have also introduced a SMBH seed at rest and
at the center of the mass of the galaxy, in order to study its mass and 
dynamical evolution.

Our simulations have been utilized fewer particles than other
works on {gas-free} disc galaxies but without an initial SMBH seed.  
In order to try to answer the
questions proposed here, we have run small simulations using the available computer 
clusters, to have, at least, an initial mass and dynamical study of the SMBH seed.

\begin{table*}
\begin{minipage}{150 mm}
\caption{Disc galaxy model properties}
\label{table1}
\begin{tabular}{@{}c|c|c|c|c|c|c|c|c|c}
Galaxy & $M_d$ & $N_d$ & $R_d$ & $Z_d$ & $R_t$ & $M_b$ & $N_b$ & $M_h$ & $N_h$\\
\hline
$G_1$ & 0.871 & 40,000 & 1.000 & 0.100 & 5.000 & 0.425 & 19,538 & 4.916 & 225,880\\
\hline
\end{tabular}

\medskip
$M_d$ is the disc mass, $N_d$ the number of particles of the disc, 
$R_d$ the disc scale radius, $Z_d$ the disc scale height,
$R_t$ the disc truncation radius,
$M_b$ the bulge mass, $N_b$ the number of particles in the bulge, 
$M_h$ the halo mass, $N_h$ the number of halo particles.
\end{minipage}
\end{table*}

\begin{table*}
\begin{minipage}{150 mm}
\caption{Continuation of Table \ref{table1}}
\label{table1a}
\begin{tabular}{@{}c|c|c|c|c}
Galaxy & $m$ & $\epsilon$ & $M_{BH}$ & $R_{bh}$\\
\hline
$G_1$ & $2.1764\times10^{-5}$ & $8.0000\times10^{-4}$ & $2.1764\times10^{-4}$ & $2.3597\times10^{-10}$\\
\hline
\end{tabular}

\medskip
$m$ the mass of each particle, and
$\epsilon$ is the softening of each particle.
$M_{bh}$ the  SMBH mass, 
$R_{bh}$ the SMBH radius, 
\end{minipage}
\end{table*}

\begin{figure}
\centering
\includegraphics[width=8cm]{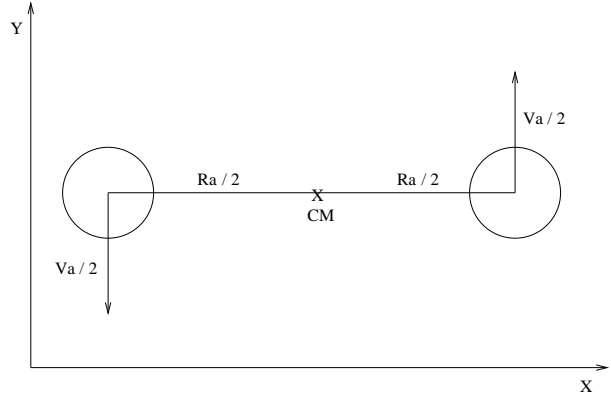}
\caption
{{Schematic plot showing the initial positions and velocities of the primary
and secondary galaxies. The quantities $R_a$ and $V_a$ are given in the
Table \ref{table3} and Table \ref{table3a}. $CM$ denotes the center
of the mass of the binary.}} 
\label{initial}
\end{figure}

\section{The results of the simulations}

\begin{figure}
\centering
\includegraphics[width=8cm]{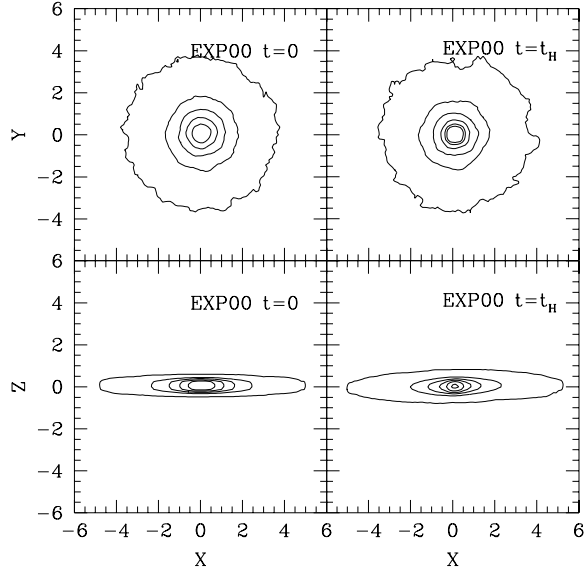}
\caption
{Contour plot of the primary galaxy $G_1$ at the times $t=0$ and $t=t_H$).
The smoothing was made by averaging the 25 first and second neighbors of
each pixel. The density levels in the planes XY and XZ at $t=0$ 
are used in contour plots, in the planes XY and XZ at $t=t_H$.}
\label{snapshot_000}
\end{figure}

\begin{figure}
\centering
\includegraphics[width=8cm]{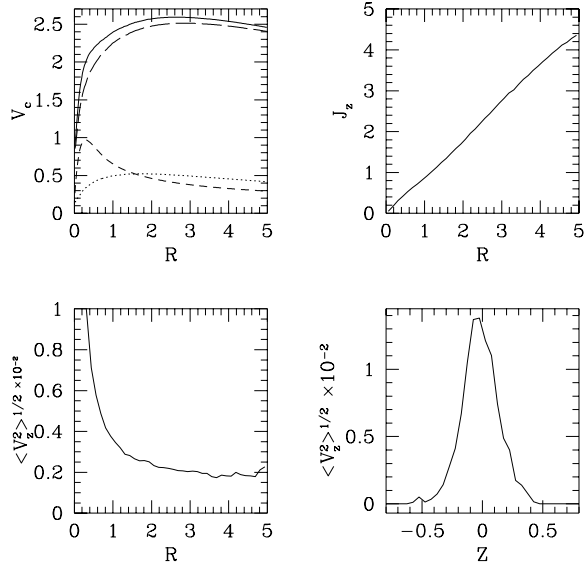}
\caption
{Rotation curve of the galaxy $G_1$ of the disc $V_c$,
the angular momentum per unit of mass $J_z$ and the
velocity dispersion in the $z$ direction $<V^2_z>^{1/2}$ at the time $t=0$.
Hereinafter, the coordinate $R$ is the  radius in cylindrical coordinates. The dotted line
denotes the disc, the long-dashed line denotes the bulge, the short-dashed
line denotes the halo, and the solid line denotes the total rotation curve.}
\label{rotcurve_000}
\end{figure}

\begin{figure}
\centering
\includegraphics[width=8cm]{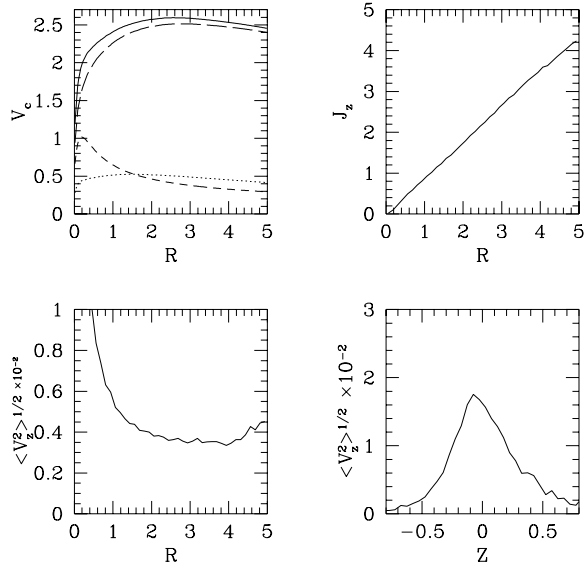}
\caption
{Rotation curve of the galaxy $G_1$ of the disc $V_c$,
the angular momentum per unit of mass $J_z$ and the
velocity dispersion in the $z$ direction $<V^2_z>^{1/2}$ at the time $t=t_H$.
The dotted line
denotes the disc, the long-dashed line denotes the bulge, the short-dashed
line denotes the halo, and the solid line denotes the total rotation curve.}
\label{rotcurve_100}
\end{figure}

\begin{figure}
\centering
\includegraphics[width=8cm]{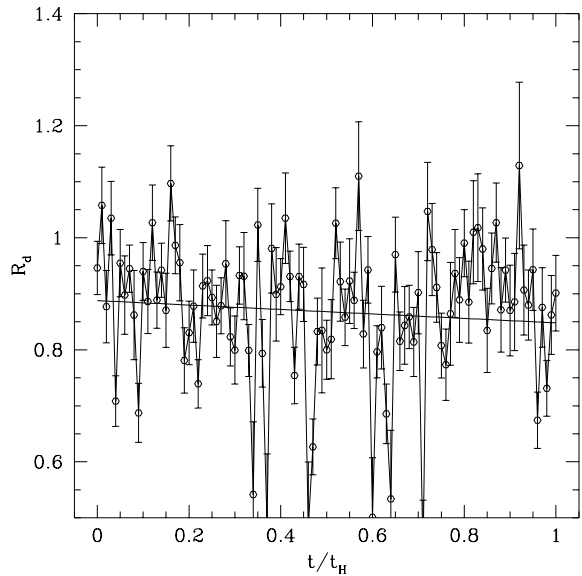}
\caption
{The evolution in time of the scale radius $R_d$.
The projected particle number density on the XY plane was fitted using the
sech disc approximation for each instant of time. 
This approximation was also used in our previous work \cite{Chan2014}.
The linear fitting parameters are $R_d=(0.8878\pm 0.1993\times 
10^{-1})[t/t_H]$ + $(0.8878\pm 0.1993\times 10^{-1})$.} 
\label{Rd}
\end{figure}

\begin{figure}
\centering
\includegraphics[width=8cm]{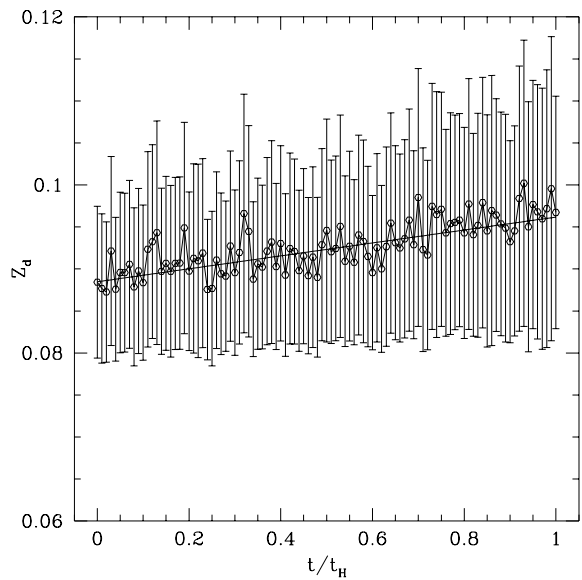}
\caption
{The evolution in time of the scale height $Z_d$.
The projected particle number density on the XZ plane was fitted using the
sech disc approximation for each instant of time, 
as used in our previous work \cite{Chan2014}.
The linear fitting parameters are $Z_d=(0.8848\times 10^{-1}\pm 0.3456\times 
10^{-3})[t/t_H]$ + $(0.7696\times 10^{-2}\pm 0.6771\times 10^{-3})$.}
\label{Zd}
\end{figure}

We have run several simulations, without the secondary galaxy to
check the initial structure of the galaxy model with the initial
SMBH seed at rest at its center (see Figures \ref{snapshot_000}-\ref{Zd}).  
For the simulations with the primary
and secondary galaxies we have used two clusters: SGI ICE-X and BULL-X BLADE B500,
The maximum number of CPU processors have used for both clusters were 32. 
Each simulation took about 45 days (BULL-X BLADE B500) and 31 days (SGI ICE-X)
of CPU time in average. 

In Figure \ref{snapshot_000} we show the contour plot of the primary 
galaxy at the beginning of the
simulation ($t=0$) and at the Hubble time of the simulation ($t=t_H$).
We note that the central density in the plane XY has increased slightly
after one
Hubble time of simulation, since the contour levels are the same for the
two moments of time.

Comparing Figures \ref{rotcurve_000} and \ref{rotcurve_100}, we note
from the quantity $<V_z^2>^{1/2}$ that
the self-heating of the initial disc and the particle halo adds
significant source of heating in the disc.  The 
softening can also cause the disc to heat up.  
Moreover, the total rotation curves $V_c$ and
the angular momentum in the Z direction have not changed at the time $t=t_H$ 
of the simulation. {The maximum of the rotation curve $V_c^{max}=2.5$ occurs at $R_{max}=2$.}

In Figures \ref{Rd} and \ref{Zd} we present the temporal evolution of the
scale radius $R_d$ and the scale height $Z_d$. 
Because of
the heating of the disc the first quantity diminishes with the time
while the second increases with the time. The linear fitting parameters
of these two quantities are presented in the captions of these figures.
In the XZ plane the scale height
has increased 8\% because of the two-body relaxation heating (Figure \ref{Zd}).

{Comparing
all the results presented in Figures \ref{snapshot_000}-\ref{Zd} 
with our previous work \cite{Chan2014} at $t=t_H$ for the same quantities
we can show that they are very
similar to ours here}, except
the rotation curve $V_c$ (Figure {\ref{rotcurve_100}) and the scale hight $Z_d$ (Figure {\ref{Zd}). {In the previous paper we have obtained
that the maximum of the rotation curve $V_c^{max}=1.2$ occurs at
$R_{max}=0.2$ and the scale hight $Z_d$ increased only $0.2\%$.}
The differences are caused mainly because of the initial SMBH seed.

All the initial conditions of the numerical experiments are presented in
Table \ref{table3} and Table \ref{table3a}. The orbits of the initial galaxies
are eccentric ($e=0.1$, $0.4$ or $0.7$) and the orientations of the discs are
coplanar ($\Theta=0$) or polar ($\Theta=90$) to each other.  
The simulations always have begun with the primary and secondary galaxies 
at the apocentric positions.

\begin{table}
\begin{minipage}{120 mm}
\caption{Primary and secondary galaxy initial conditions}
\label{table3}
\begin{tabular}{@{}c|c|c|c|c|c|c}
EXP& $\Theta$ & $R_p$ & $e$ &    $R_a$  &   $V_a$ &       $M_{bh}$      \\
\hline
00 &         &     &        &           &         & $2.1764\times10^{-4}$\\
\hline
01 &     0   &  12 &    0.1 &    14.67  &  0.8732 & $2.1764\times10^{-4}$\\
\hline
02 &     0   &  12 &    0.4 &    28.00  &  0.5160 & $2.1764\times10^{-4}$\\
\hline
03 &     0   &  12 &    0.7 &    68.00  &  0.2341 & $2.1764\times10^{-4}$\\
\hline
04 &     0   &  15 &    0.1 &    18.33  &  0.7810 & $2.1764\times10^{-4}$\\
\hline
05 &     0   &  15 &    0.4 &    35.00  &  0.4615 & $2.1764\times10^{-4}$\\
\hline
06 &     0   &  15 &    0.7 &    85.00  &  0.2094 & $2.1764\times10^{-4}$\\
\hline
07 &     0   &  20 &    0.1 &    24.44  &  0.6763 & $2.1764\times10^{-4}$\\
\hline
08 &     0   &  20 &    0.4 &    46.67  &  0.3997 & $2.1764\times10^{-4}$\\
\hline
09 &     0   &  20 &    0.7 &   113.33  &  0.1814 & $2.1764\times10^{-4}$\\
\hline
10 &     0   &  23 &    0.1 &    28.11  &  0.6307 & $2.1764\times10^{-4}$\\
\hline
11 &     0   &  23 &    0.4 &    53.67  &  0.3727 & $2.1764\times10^{-4}$\\
\hline
12 &     0   &  23 &    0.7 &   130.33  &  0.1691 & $2.1764\times10^{-4}$\\
\hline
13 &     0   &  25 &    0.1 &    30.56  &  0.6049 & $2.1764\times10^{-4}$\\
\hline
14 &     0   &  25 &    0.4 &    58.33  &  0.3575 & $2.1764\times10^{-4}$\\
\hline
15 &     0   &  25 &    0.7 &   141.67  &  0.1622 & $2.1764\times10^{-4}$\\
\hline
16 &     0   &  30 &    0.1 &    36.67  &  0.5522 & $2.1764\times10^{-4}$\\
\hline
17 &     0   &  30 &    0.4 &    70.00  &  0.3263 & $2.1764\times10^{-4}$\\
\hline
18 &     0   &  30 &    0.7 &   170.00  &  0.1481 & $2.1764\times10^{-4}$\\
\hline
\end{tabular}

\medskip
$\Theta$ the angle between the two planes of the discs in units of degree,
$R_p$ the pericentric distance, $M_1$ the primary galaxy mass, 
$e$ the eccentricity, $R_a$ the apocentric distance,
$V_a$ the velocity at the apocentric distance, 
$M_1$ the primary galaxy mass, and $M_2=M_1=0.621$ is the secondary mass galaxy. 
In these experiments the orbits of the particles of both galaxies, primary and
secondary galaxy, have clockwise rotations.
\end{minipage}
\end{table}

\begin{table}
\begin{minipage}{120 mm}
\caption{Continuation of Table \ref{table3}}
\label{table3a}
\begin{tabular}{@{}c|c|c|c|c|c|c}
EXP& $\Theta$ & $R_p$ & $e$ &    $R_a$  &   $V_a$ &       $M_{bh}$      \\
\hline
19 &    90   &  12 &    0.1 &    14.67  &  0.8732 & $2.1764\times10^{-4}$\\
\hline
20 &    90   &  12 &    0.4 &    28.00  &  0.5160 & $2.1764\times10^{-4}$\\
\hline
21 &    90   &  12 &    0.7 &    68.00  &  0.2341 & $2.1764\times10^{-4}$\\
\hline
22 &    90   &  15 &    0.1 &    18.33  &  0.7810 & $2.1764\times10^{-4}$\\
\hline
23 &    90   &  15 &    0.4 &    35.00  &  0.4615 & $2.1764\times10^{-4}$\\
\hline
24 &    90   &  15 &    0.7 &    85.00  &  0.2094 & $2.1764\times10^{-4}$\\
\hline
25 &    90   &  20 &    0.1 &    24.44  &  0.6763 & $2.1764\times10^{-4}$\\
\hline
26 &    90   &  20 &    0.4 &    46.67  &  0.3997 & $2.1764\times10^{-4}$\\
\hline
27 &    90   &  20 &    0.7 &   113.33  &  0.1814 & $2.1764\times10^{-4}$\\
\hline
28 &    90   &  23 &    0.1 &    28.11  &  0.6307 & $2.1764\times10^{-4}$\\
\hline
29 &    90   &  23 &    0.4 &    53.67  &  0.3727 & $2.1764\times10^{-4}$\\
\hline
30 &    90   &  23 &    0.7 &   130.33  &  0.1691 & $2.1764\times10^{-4}$\\
\hline
31 &    90   &  25 &    0.1 &    30.56  &  0.6049 & $2.1764\times10^{-4}$\\
\hline
32 &    90   &  25 &    0.4 &    58.33  &  0.3575 & $2.1764\times10^{-4}$\\
\hline
33 &    90   &  25 &    0.7 &   141.67  &  0.1622 & $2.1764\times10^{-4}$\\
\hline
34 &    90   &  30 &    0.1 &    36.67  &  0.5522 & $2.1764\times10^{-4}$\\
\hline
35 &    90   &  30 &    0.4 &    70.00  &  0.3263 & $2.1764\times10^{-4}$\\
\hline
36 &    90   &  30 &    0.7 &   170.00  &  0.1481 & $2.1764\times10^{-4}$\\
\hline
\end{tabular}

\medskip
$\Theta$ the angle between the two planes of the discs in units of degree,
$R_p$ the pericentric distance, $M_1$ the primary galaxy mass, 
$e$ the eccentricity, $R_a$ the apocentric distance,
$V_a$ the velocity at the apocentric distance, 
$M_1$ the primary galaxy mass, and $M_2=M_1=0.621$ is the secondary mass 
galaxy. 
In these experiments the orbits of the particles of both galaxies, primary and
secondary galaxy, have clockwise rotations.
\end{minipage}
\end{table}

We will show only the evolution time of the SMBH of the experiments where the two discs
merge or graze each other, where the tidal effects are more proeminent during the evolution of
the simulation (see Table \ref{table5}-\ref{table5a}). 
The experiments are: EXP02 (Figure \ref{part_accret_02_bh1a}), 
EXP06 (Figure \ref{part_accret_06_bh1}), 
EXP20 (Figure \ref{part_accret_20_bh1}) and EXP24 (Figure \ref{part_accret_24_bh1}).

Comparing the contour plots of the discs at $t=t_H$ shown in Figure \ref{ccontorxyz_02_20_bh1} 
for the experiments EXP02 and EXP20 we can note that the 
merger of the primary and secondary discs can result in a final normal disc (EXP02) or a 
final warped disc (EXP20).
After the fusion of discs, the final one is thicker and larger than the initial discs.

\begin{figure}
\centering
\includegraphics[width=8cm]{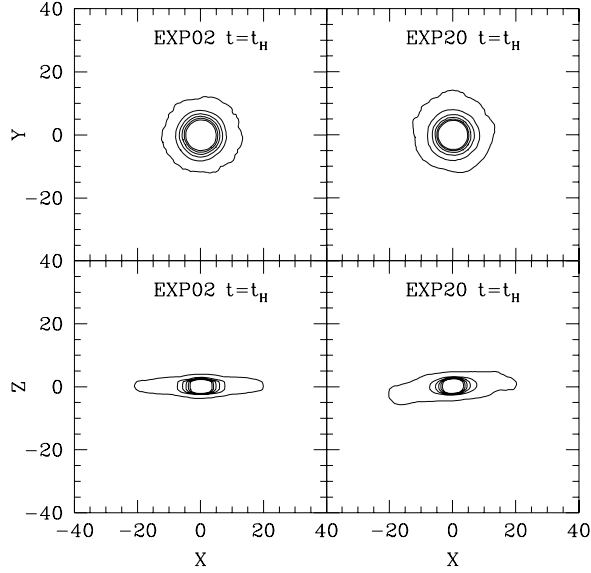}
\caption
{Contour plots of the final merged discs at $t=t_H$ of the experiments EXP02 and EXP20.}
\label{ccontorxyz_02_20_bh1}
\end{figure}

In Figures \ref{part_accret_02_bh1a}a-\ref{part_accret_24_bh1}a we show
the time evolution of the SMBH mass of the primary and secondary galaxy of the experiments 
EXP02, EXP06, EXP20 and EXP24. 
We also present the time evolution 
of the SMBH mass of the isolated galaxy in the 
Figures \ref{part_accret_02_bh1a}a-\ref{part_accret_24_bh1}a, 
in order to compare its SMBH mass growth
to the SMBH mass growth of the primary and secondary galaxy during the evolution. 
In the same plots we show the temporal
evolution of the distance of the center of mass between the two galaxies. 
There is an arbitrary scale factor only to adjust the distance 
within the plot scale of each figure. The purpose of these plots is to know if the distance
approach of the primary and secondary galaxies to each other increases the mass growth of the SMBHs. 
 
The Figures \ref{part_accret_02_bh1a}b-\ref{part_accret_24_bh1}b and
Figures \ref{part_accret_02_bh1a}c-\ref{part_accret_24_bh1}c show
the time evolution of the number of accreted particles of the primary and 
secondary galaxy onto the SMBHs, respectively. 
Theses plots show how many halo, bulge and disc particles that contribute to the growth of SMBH mass. 

Comparing the Figures \ref{part_accret_02_bh1a}a-\ref{part_accret_24_bh1}a we can notice
the tidal effects in the SMBH mass of the secondary galaxy are more important 
(see the distance of the center of mass between the two galaxies in the plot). The approach of the
galaxies to each other seems not to affect too much the primary galaxy (see Table \ref{table5}-\ref{table5a}).

{
Comparing the  
Figures \ref{part_accret_02_bh1a}b-\ref{part_accret_24_bh1}b and Figures 
\ref{part_accret_02_bh1a}c-\ref{part_accret_24_bh1}c, respectively, 
we can note that 
most of the accreted particles {onto} the SMBH have come from the bulges and from the halos. 
Only a small number of the accreted particles {onto} the SMBH has come from the discs.
}

In some cases of final merging stage of the two galaxies, 
the final SMBH of the secondary galaxy is {ejected out of the galaxy} (see 
Table \ref{table5}-\ref{table5a}). 

\begin{figure}
\centering
\includegraphics[width=8cm]{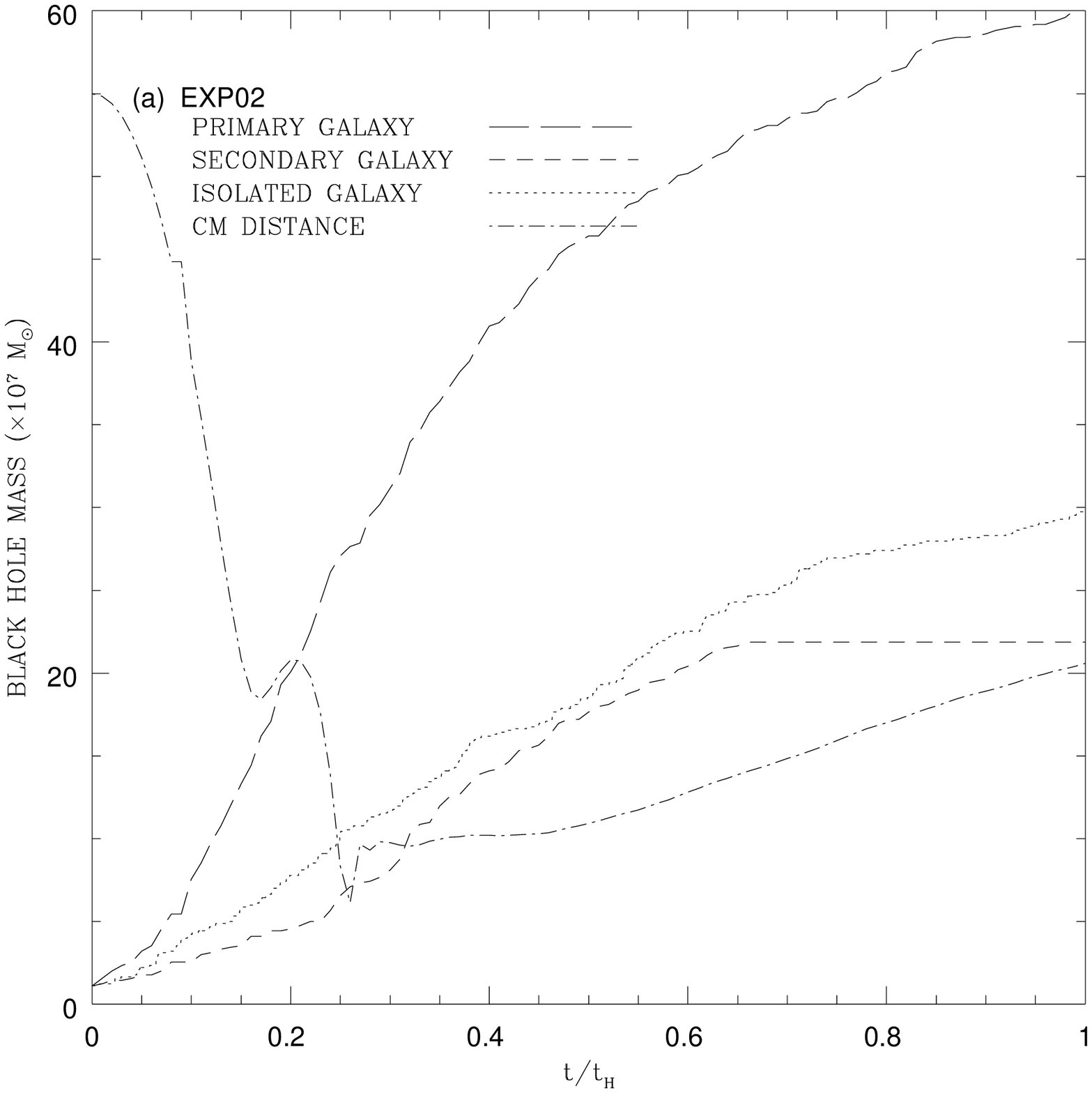}
\includegraphics[width=8cm]{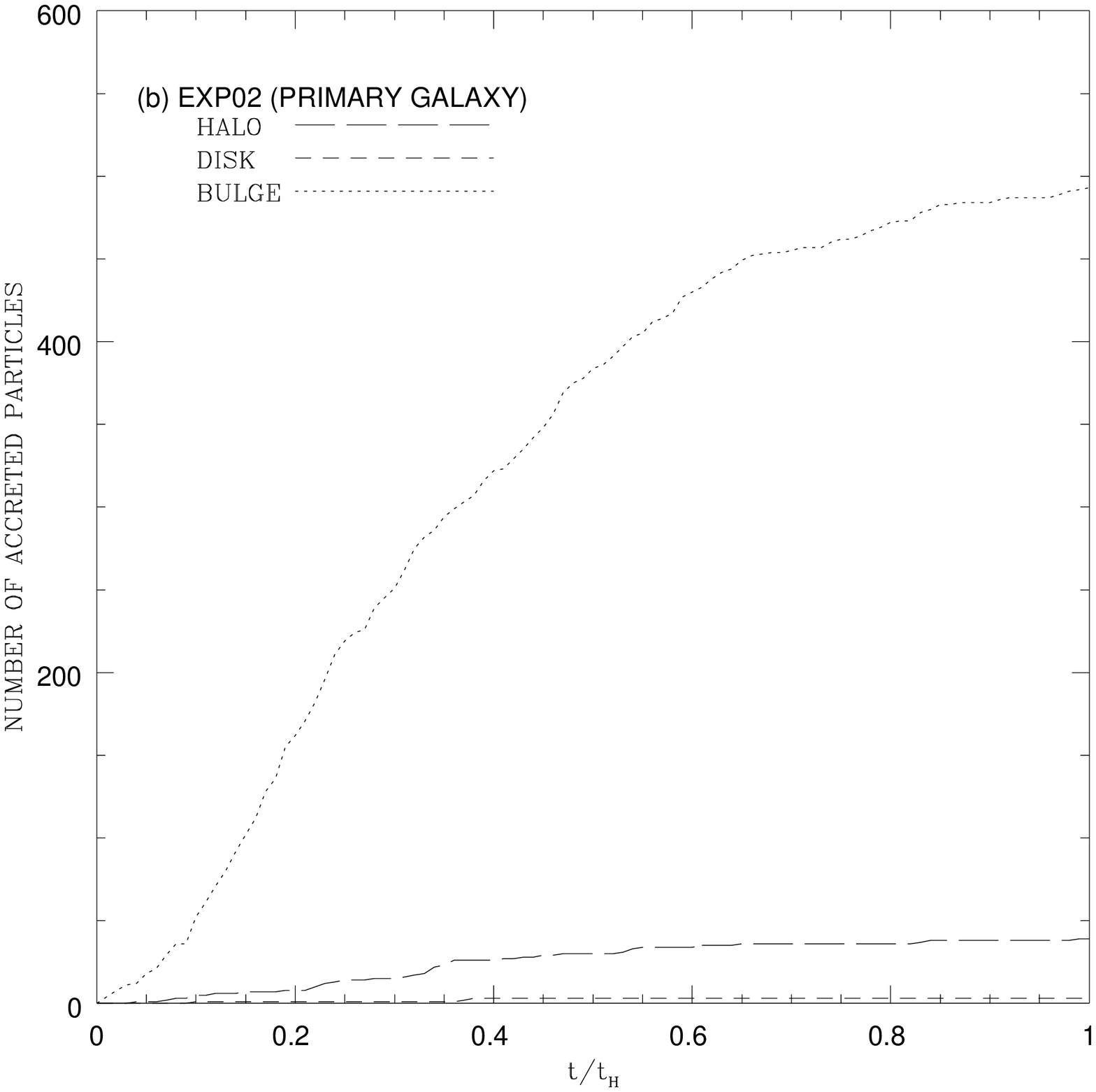}
\includegraphics[width=8cm]{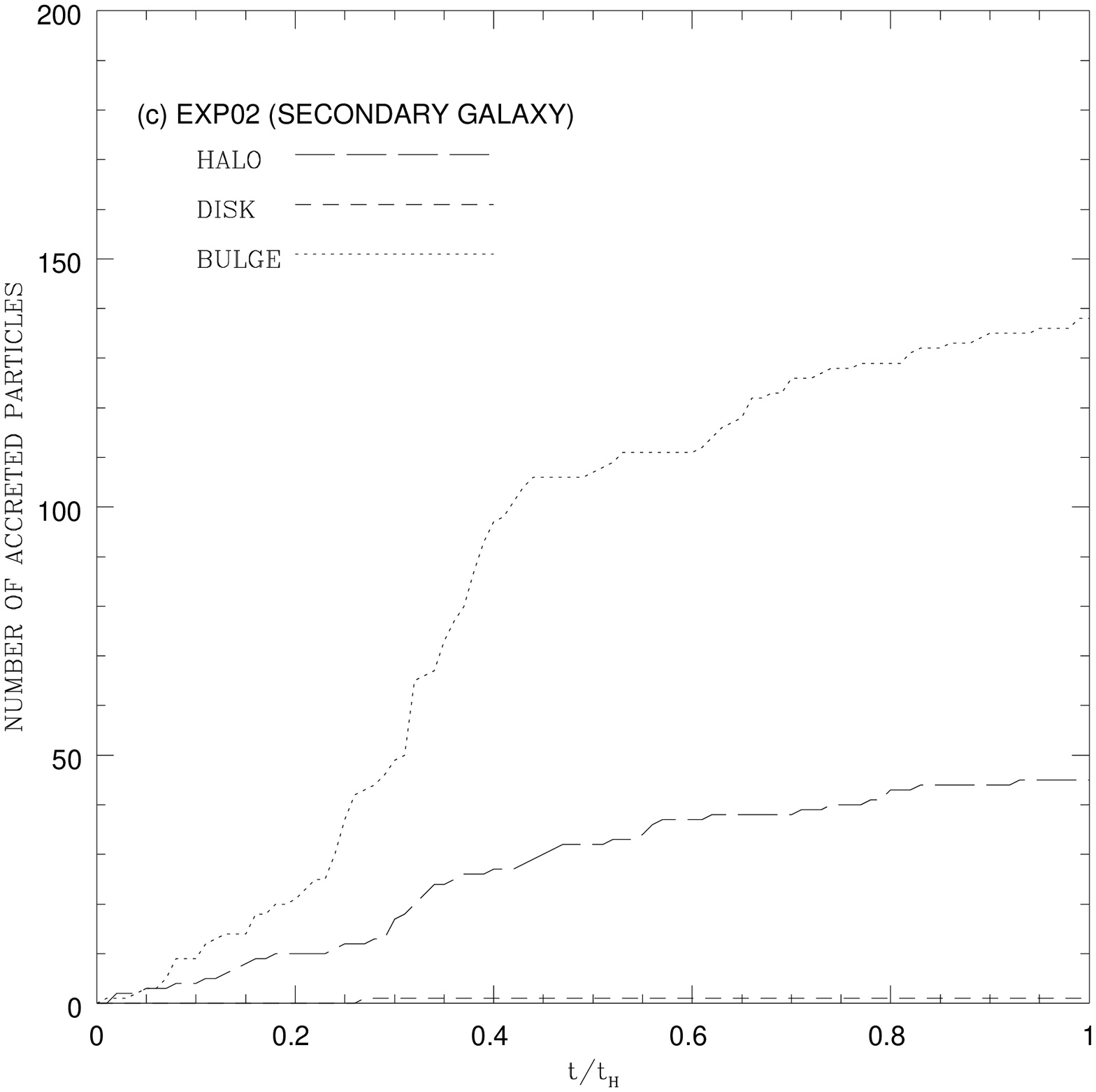}
\caption
{(a) Temporal evolution of the SMBH seed mass of the primary (long-dashed line) 
and secondary galaxy (short-dashed line) of the experiment EXP02. 
We also present the time evolution 
of the SMBH seed mass of the isolated galaxy. In the same plot we show the temporal
evolution of the distance of the center of mass of the two galaxies 
(dot-dashed line). There is an arbitrary scale factor only to adjust 
the distance within the plot scale. (b) and (c) Time evolution of number of accreted
particles of the primary and secondary galaxy {onto} the SMBH. The long-dashed lines 
represent the halo particles. 
The dotted lines represent the bulge particles. The short-dashed lines 
represent the disk particles.}
\label{part_accret_02_bh1a}
\end{figure}

\begin{figure}
\centering
\includegraphics[width=8cm]{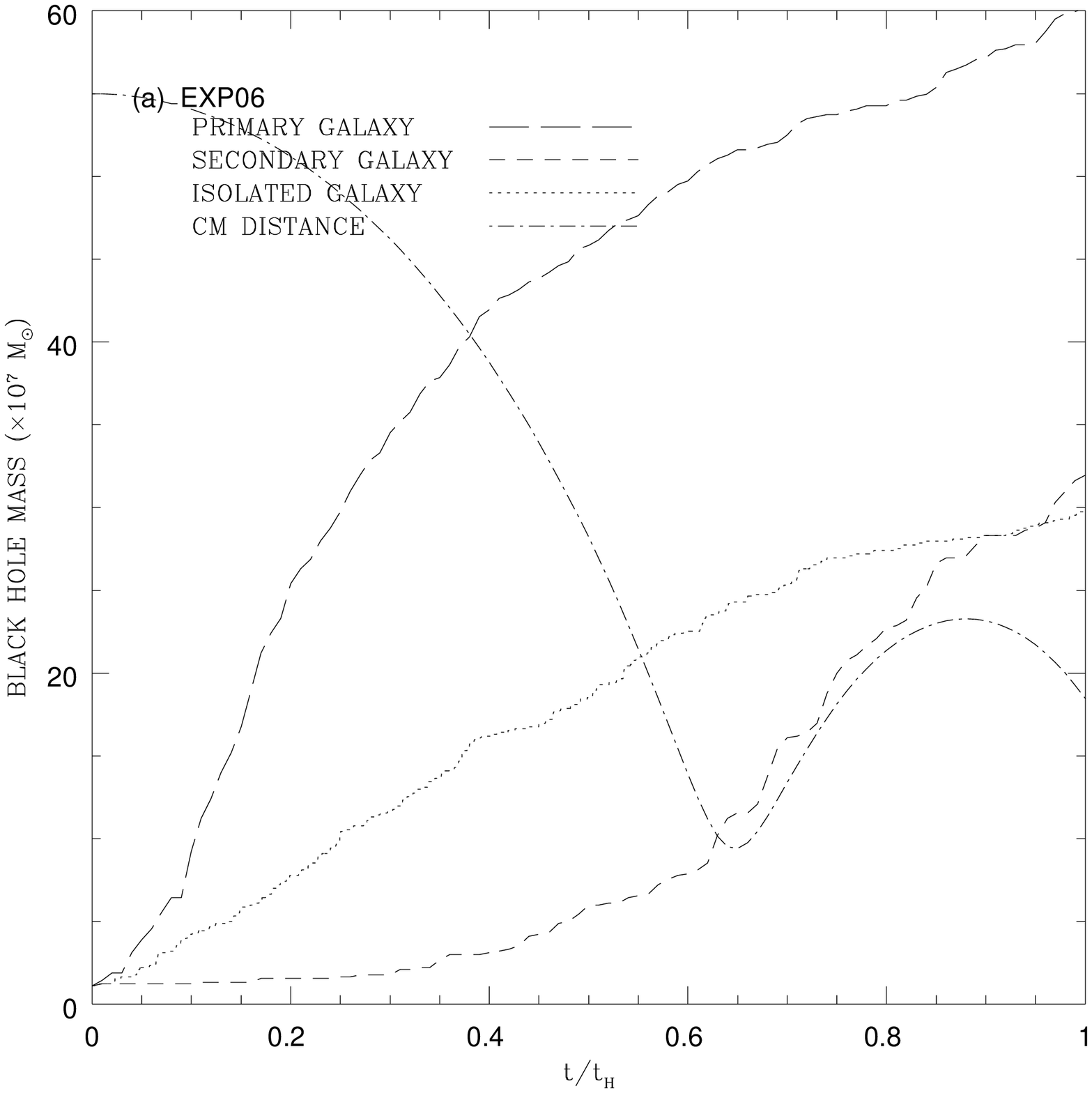}
\includegraphics[width=8cm]{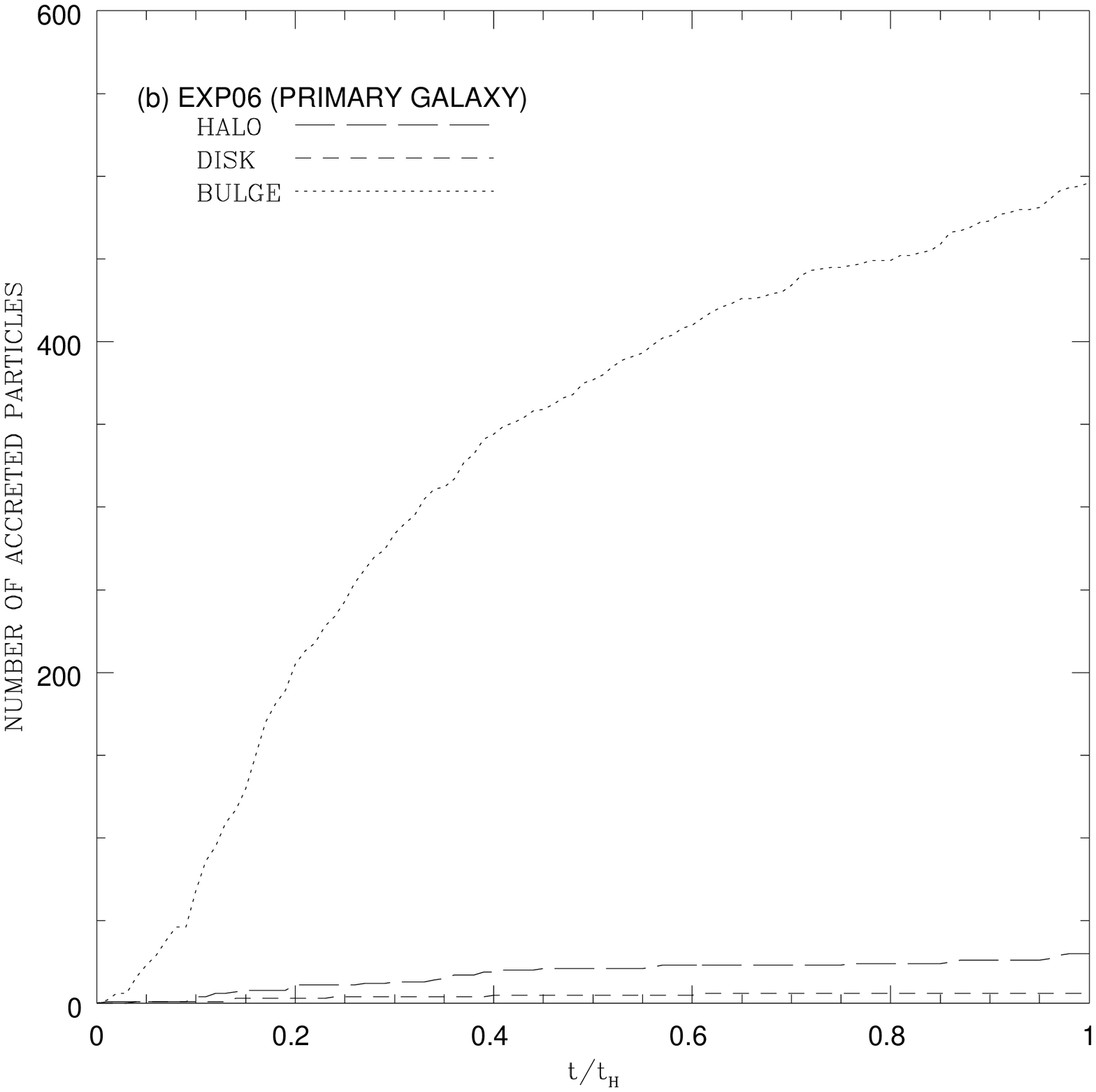}
\includegraphics[width=8cm]{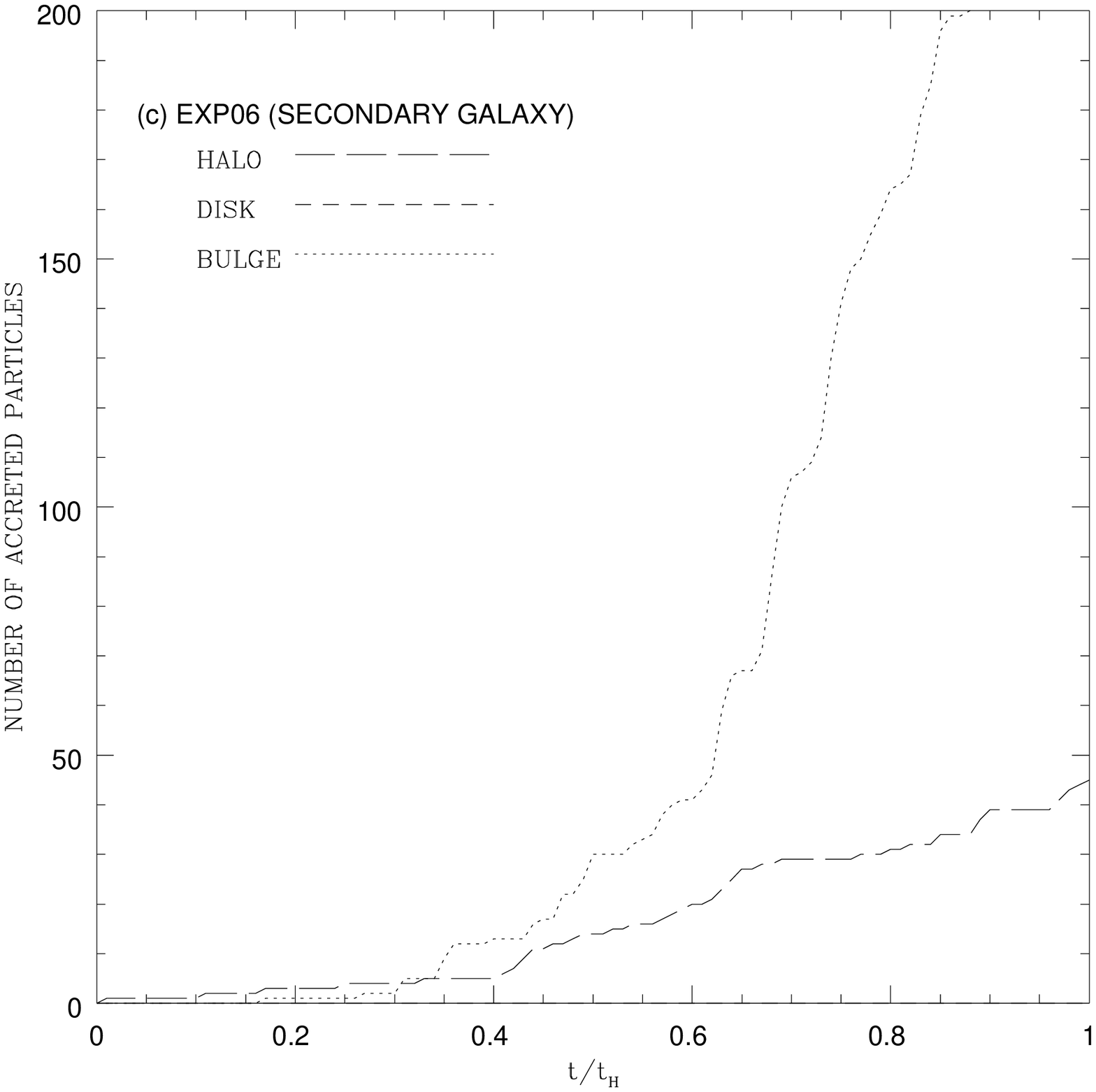}
\caption
{(a) Temporal evolution of the SMBH seed mass of the primary (long-dashed line) 
and secondary galaxy (short-dashed line) of the experiment EXP06. 
{We also present the time evolution 
of the SMBH seed mass of the isolated galaxy. In the same plot we show the temporal
evolution of the distance of the center of mass of the two galaxies 
(dot-dashed line). There is an arbitrary scale factor only to adjust 
the distance within the plot scale. (b) and (c) Time evolution of number of accreted
particles of the primary and secondary galaxy {onto} the SMBH. The long-dashed lines 
represent the halo particles. 
The dotted lines represent the bulge particles. The short-dashed lines 
represent the disk particles}.}
\label{part_accret_06_bh1}
\end{figure}

\begin{figure}
\centering
\includegraphics[width=8cm]{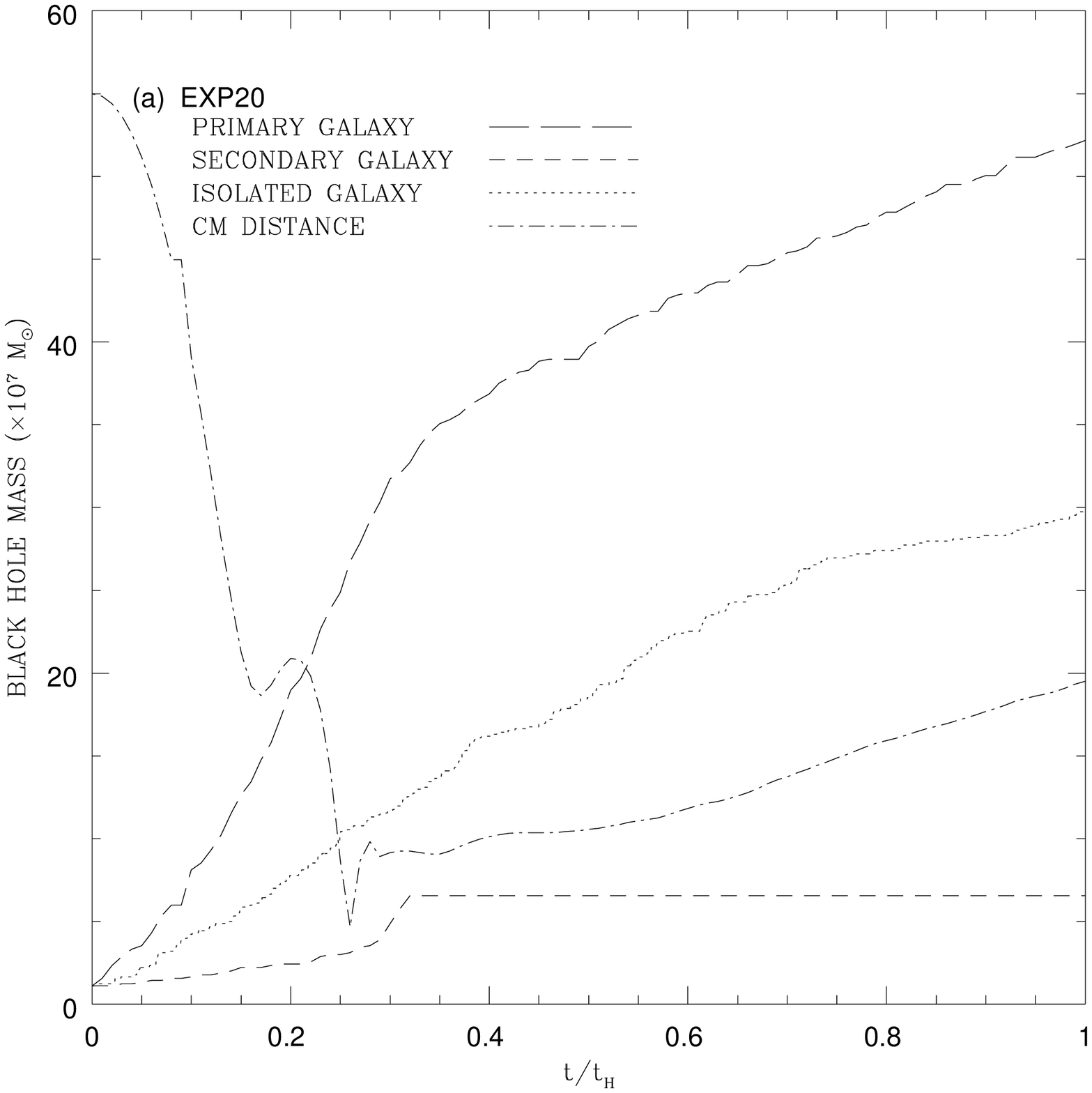}
\includegraphics[width=8cm]{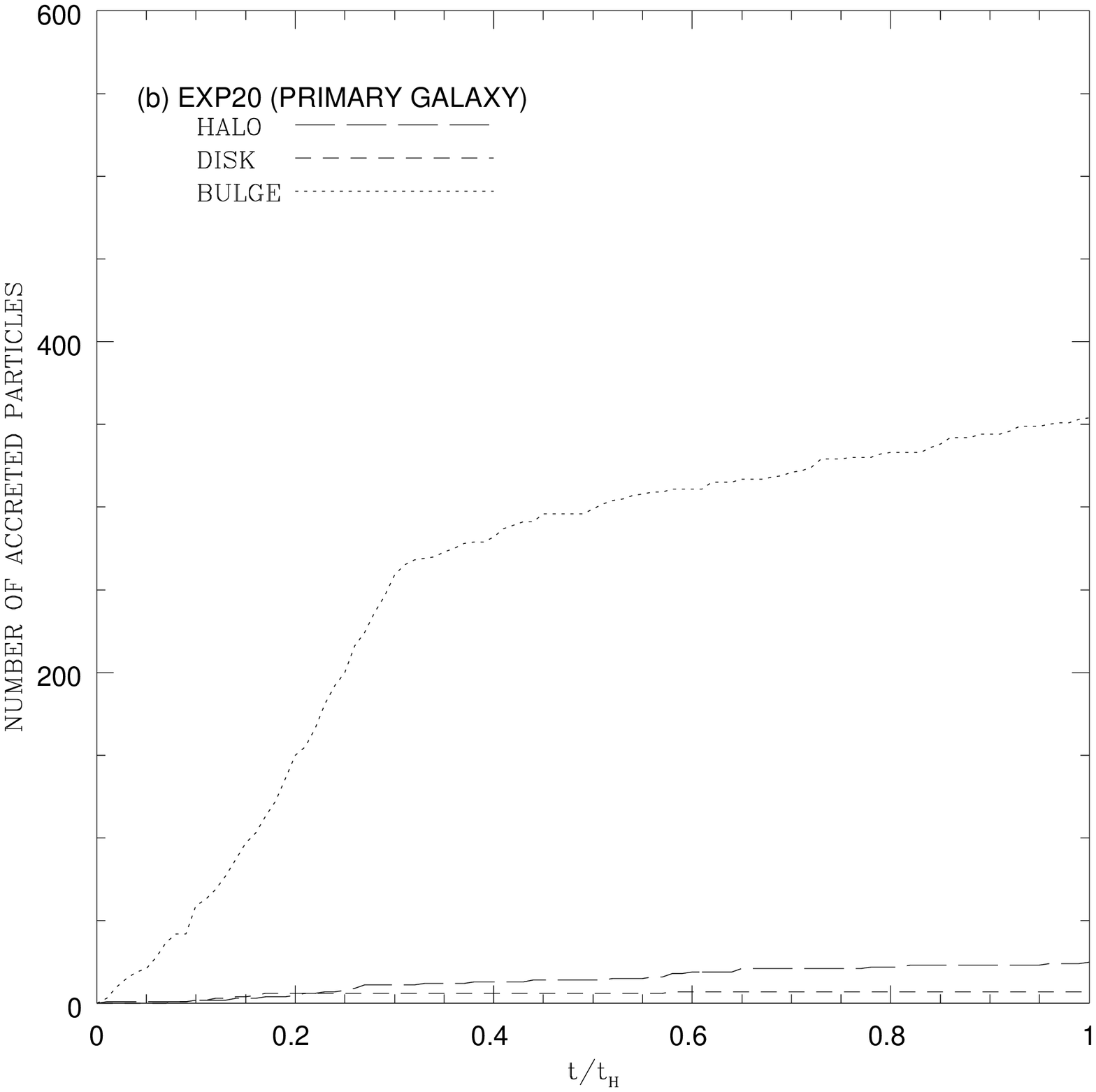}
\includegraphics[width=8cm]{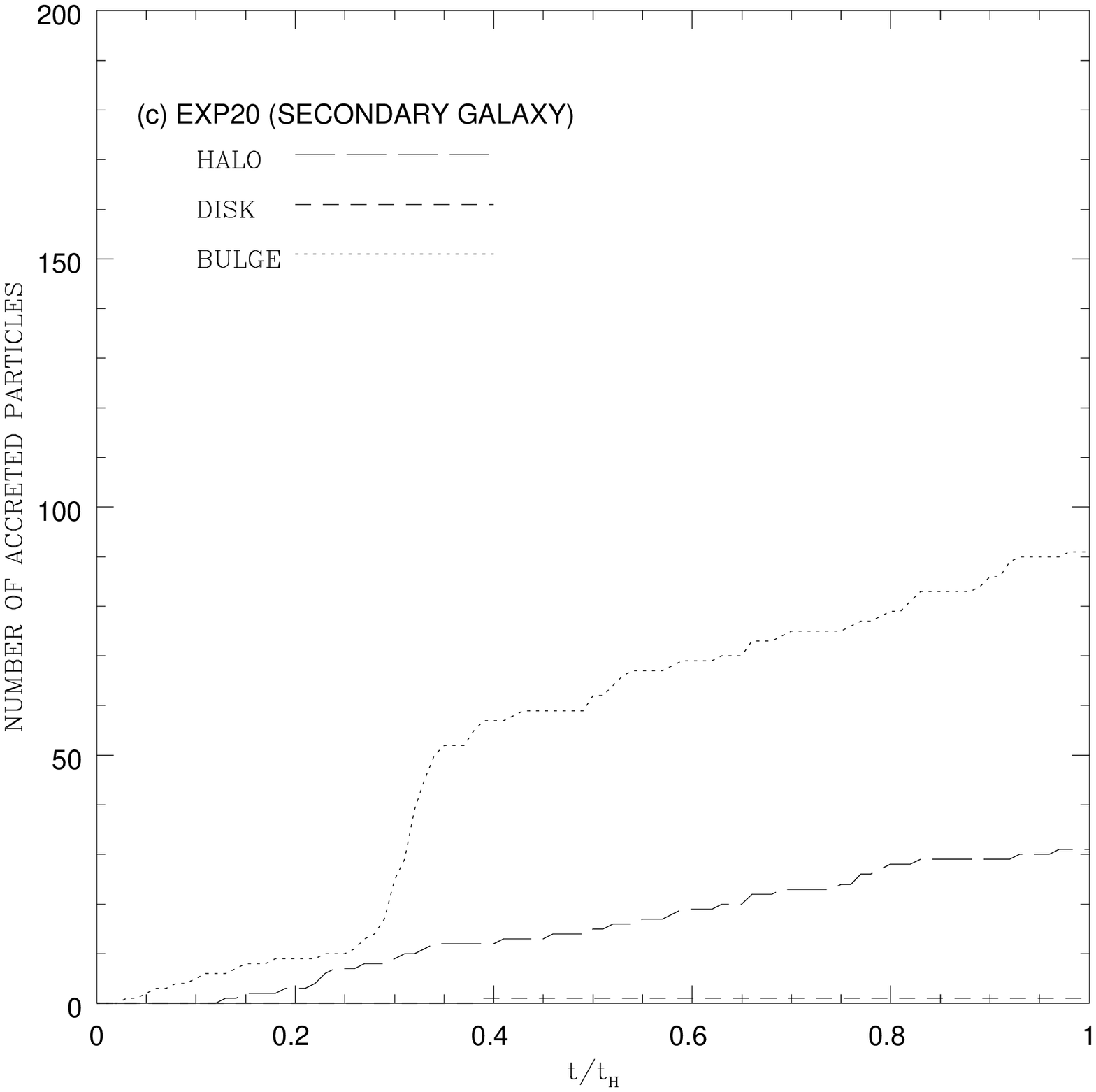}
\caption
{(a) Temporal evolution of the SMBH seed mass of the primary (long-dashed line) 
and secondary galaxy (short-dashed line) of the experiment EXP20. 
{We also present the time evolution 
of the SMBH seed mass of the isolated galaxy. In the same plot we show the temporal
evolution of the distance of the center of mass of the two galaxies 
(dot-dashed line). There is an arbitrary scale factor only to adjust 
the distance within the plot scale. (b) and (c) Time evolution of number of accreted
particles of the primary and secondary galaxy {onto} the SMBH. The long-dashed lines 
represent the halo particles. 
The dotted lines represent the bulge particles. The short-dashed lines 
represent the disk particles}.}
\label{part_accret_20_bh1}
\end{figure}

\begin{figure}
\centering
\includegraphics[width=8cm]{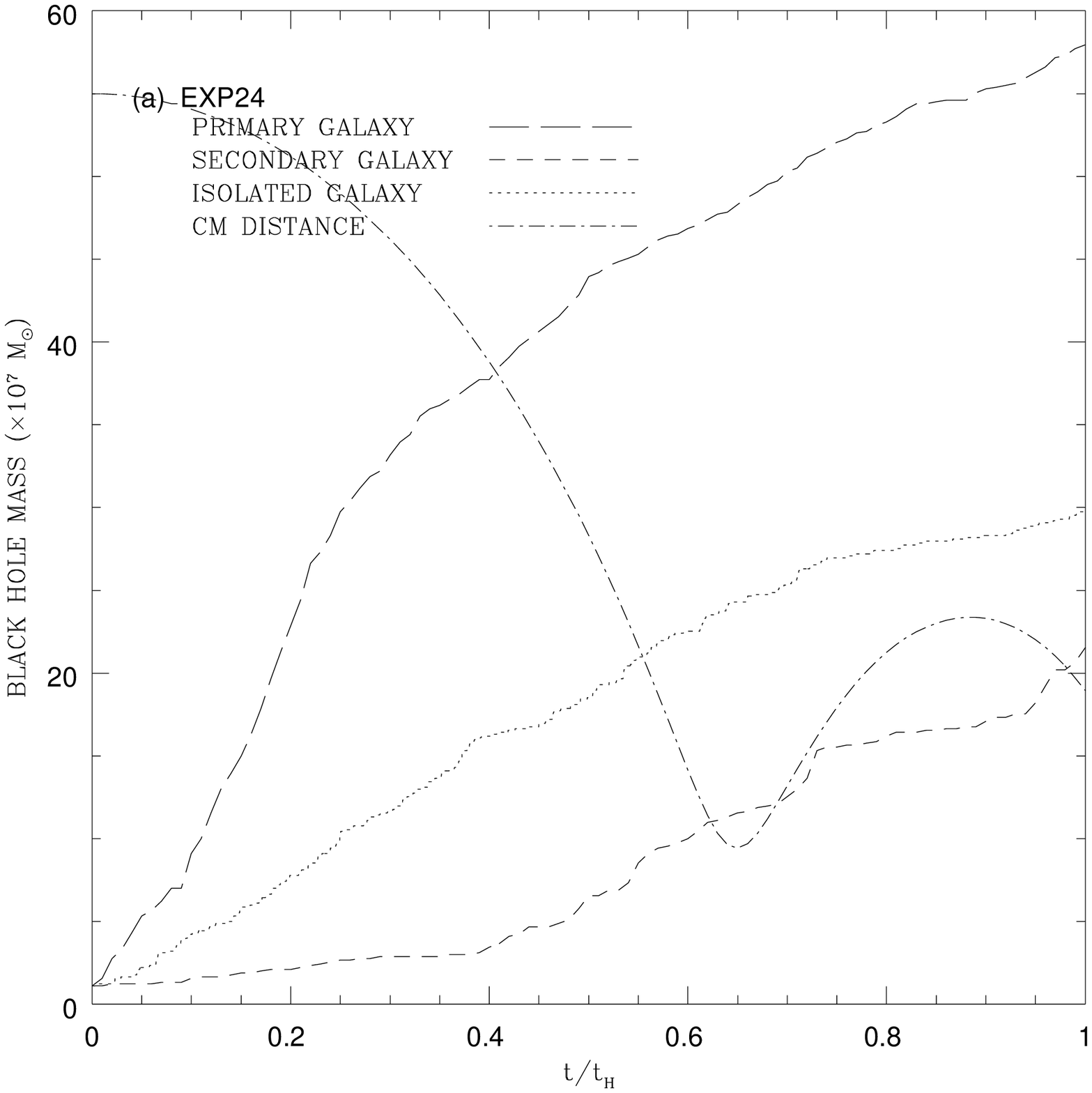}
\includegraphics[width=8cm]{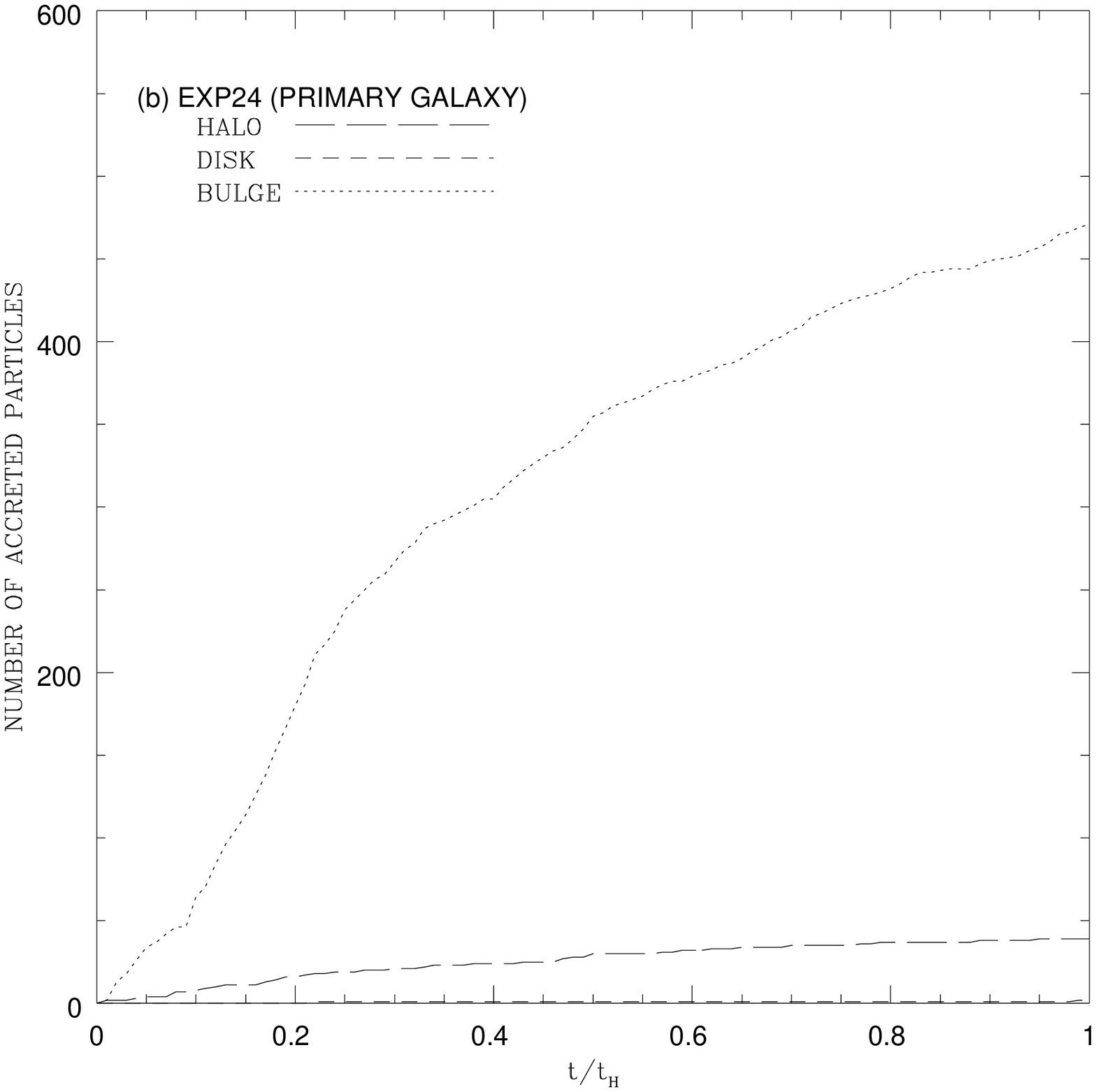}
\includegraphics[width=8cm]{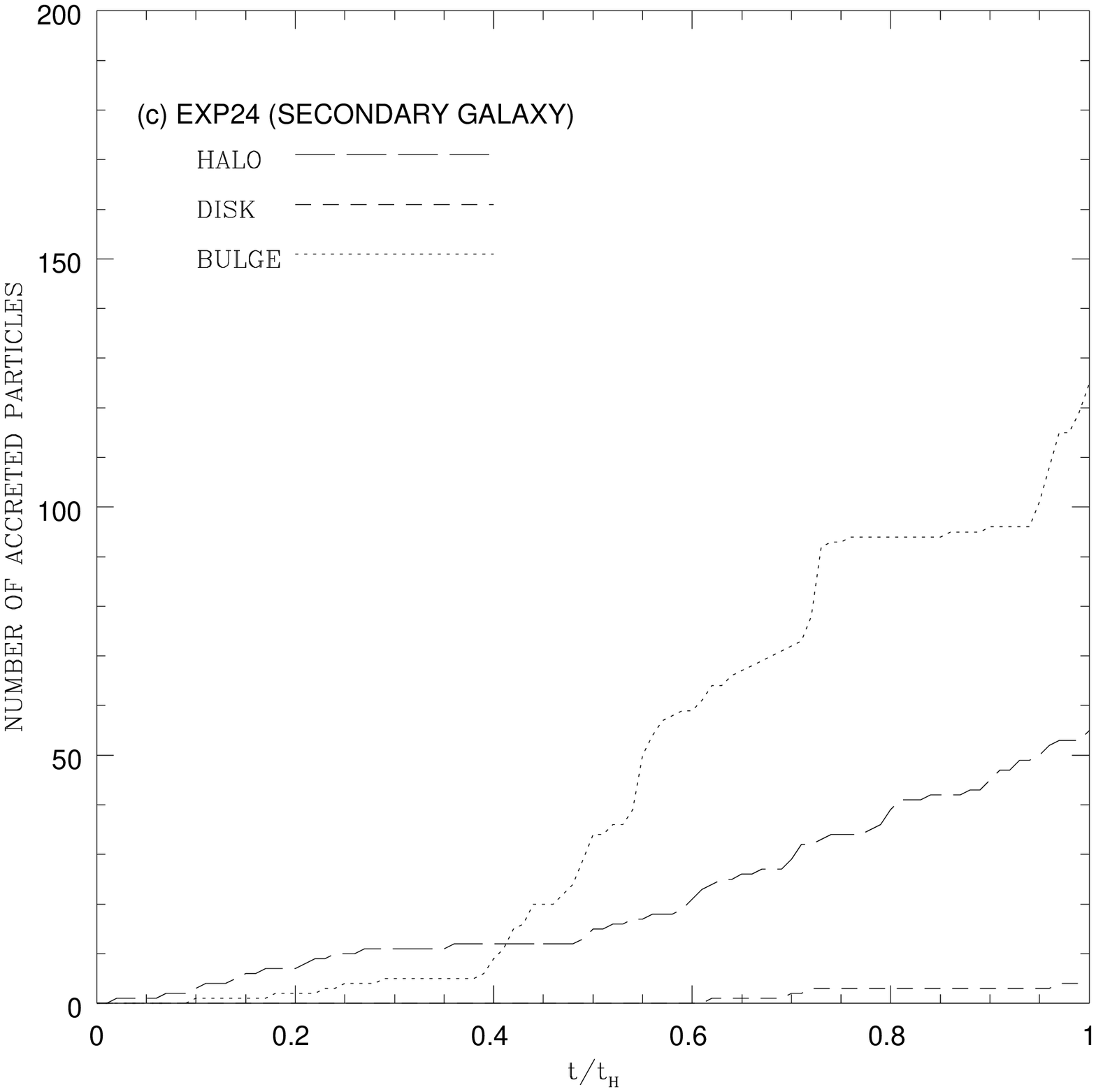}
\caption
{(a) Temporal evolution of the SMBH seed mass of the primary (long-dashed line) 
and secondary galaxy (short-dashed line) of the experiment EXP24. 
{We also present the time evolution 
of the SMBH seed mass of the isolated galaxy. In the same plot we show the temporal
evolution of the distance of the center of mass of the two galaxies 
(dot-dashed line). There is an arbitrary scale factor only to adjust 
the distance within the plot scale. (b) and (c) Time evolution of number of accreted
particles of the primary and secondary galaxy {onto} the SMBH. The long-dashed lines 
represent the halo particles. 
The dotted lines represent the bulge particles. The short-dashed lines 
represent the disk particles}}.
\label{part_accret_24_bh1}
\end{figure}

From Table \ref{table5}-\ref{table5a} we can see comparing the final SMBH mass of all the 
experiments that the mass of the SMBH of the primary galaxy have increased by a factor ranging 
from 52 to 64 times the initial seed mass, depending on the experiment. However,
the mass of the SMBH of the secondary galaxy have increased by a factor ranging 
from 6 to 33 times in comparison to the initial seed mass, depending on the experiment. 
Thus, we can conclude that the tidal effects are very important, modifying the evolution of the SMBH 
in the primary and secondary galaxy differently.

\begin{table}
\begin{minipage}{150 mm}
\caption{Characteristics of galaxy orbits and SMBH mass at $t_H$}
\label{table5}
\begin{tabular}{@{}c|c|c|c}
EXP & Disc Interaction & Primary SMBH  & Secondary SMBH \\
\hline
01 &  Merge  & 58.0533 & 16.7611* \\
\hline
02 &  Merge  & 60.1646 & 21.8667* \\
\hline
03 &  Merge  & 63.4899 & 23.3100* \\
\hline
04 &  Merge  & 52.2800 & 33.8553 \\
\hline
05 &  Merge  & 58.4970 & 14.4299* \\
\hline
06 &  Graze  & 60.1646 & 31.9683 \\
\hline
07 &  Merge  & 59.1651 & 29.4152 \\
\hline
08 &  Merge  & 67.0445 & 31.9678 \\
\hline
09 & Distant & 57.4974 & 17.3160 \\
\hline
10 &  Merge  & 59.4966 & 30.7468 \\
\hline
11 & Distant & 58.9406 & 23.8649 \\
\hline
12 & Distant & 62.9391 & 16.4281 \\
\hline
13 &  Merge  & 60.7155 & 31.3022 \\
\hline
14 & Distant & 61.9394 & 22.9770 \\
\hline
15 & Distant & 61.3836 & 18.6481 \\
\hline
16 & Distant & 58.7214 & 27.3059 \\
\hline
17 & Distant & 64.2702 & 21.3118 \\
\hline
18 & Distant & 60.2718 & 15.4290 \\
\hline
\end{tabular}

\medskip
Initial mass of SMBH of the primary and secondary galaxy is $1.1099$,
Primary SMBH and Secondary SMBH in units of $10^7 M_\odot$,
$Graze$ means that the two discs touch each other for a 
while and then separate.  $Merge$ means that the two discs fuse. $Distant$ means
the two discs interacts apart each other. The symbol (*) after the mass of the 
SMBH of some merging experiments means that the SMBH has been {ejected out of the}
binary system during the evolution of the merged binary.

\end{minipage}
\end{table}

\begin{table}
\begin{minipage}{150 mm}
\caption{Continuation of Table \ref{table5}}
\label{table5a}
\begin{tabular}{@{}c|c|c|c}
EXP & Disc Interaction & Primary SMBH & Secondary SMBH \\
\hline
19 &  Merge  & 53.6111 & 26.8622 \\
\hline
20 &  Merge  & 52.1679 &  6.5489* \\
\hline
21 &  Merge  & 62.0517 & 16.5387* \\
\hline
22 &  Merge  & 53.3919 & 27.3059 \\
\hline
23 &  Merge  & 54.1670 & 13.8750 \\
\hline
24 &  Graze  & 57.9411 & 21.5337 \\
\hline
25 &  Merge  & 54.8352 & 14.2080 \\
\hline
26 &  Merge  & 62.0517 & 21.8667 \\
\hline
27 & Distant & 58.1654 & 21.2012 \\
\hline
28 &  Merge  & 58.8284 & 23.4212 \\
\hline
29 & Distant & 60.4962 & 24.3091 \\
\hline
30 & Distant & 61.3836 & 18.8700 \\
\hline
31 &  Merge  & 61.8273 & 21.3118 \\
\hline
32 & Distant & 62.3832 & 21.3118 \\
\hline
33 & Distant & 61.7151 & 21.7560 \\
34 & Distant & 61.6029 & 24.8640 \\
\hline
35 & Distant & 59.3843 & 21.9779 \\
\hline
36 & Distant & 61.0521 & 21.2007 \\
\hline
\end{tabular}

\medskip
Initial mass of SMBH of the primary and secondary galaxy is $1.1099$,
Primary SMBH and Secondary SMBH in units of $10^7 M_\odot$,
$Graze$ means that the two discs touch each other for a 
while and then separate.  $Merge$ means that the two discs fuse. $Distant$ means
the two discs interacts apart each other. The symbol (*) after the mass of the 
SMBH of some merging experiments means that the SMBH has been {ejected out of the}
binary system during the evolution of the merged binary.

\end{minipage}
\end{table}

\section{Conclusions}
~We have shown the results of N-body simulations of the 
interactions of two gas-free disc galaxies with the same mass. Both disc galaxies have 
halos of dark matter, central bulges and initial SMBH
seeds at their centers. 

We have found that the 
merger of the primary and secondary discs can result in a final normal disc 
or a final warped disc. After the fusion of discs,
the final one is thicker and larger than the initial disc.

The tidal effects are very important, modifying the evolution of the SMBH 
in the primary and secondary galaxy differently.  
The mass of the SMBH of the primary galaxy have increased by a factor ranging 
from 52 to 64 times 
the initial seed mass, depending on the experiment. However,
the mass of the SMBH of the secondary galaxy have increased by a factor ranging 
from 6 to 33 times the initial SMBH seed mass, depending also on the experiment. 

Most of the accreted particles have come from the bulges and from the halos, depleting their particles. This could explain why the observations show
that the SMBH with masses of approximately $10^6 M_\odot$ are found in
many bulgeless galaxies \cite{Kormendy2013}. However,
only a small number of the accreted particles has come from the disc.

In some cases of final merging stage of the two galaxies, 
the final SMBH of the secondary galaxy was {ejected out of the galaxy}. 

\bigskip
\noindent {\bf ACKNOWLEDGMENTS}
\bigskip

The author
acknowledges the the financial support
from Conselho Nacional de Desenvolvimento Cient\'{\i}fico e Tecnol\'ogico -
Brazil.

The author also thanks the generous amount of CPU time given by 
CENAPAD/UFC (Centro Nacional de Processamento de Alto Desempenho da UFC) and
NACAD/COPPE-UFRJ (N\'ucleo de Avan\c cado de Computa\c c\~ao de Alto
Desempenho da COPPE/UFRJ) in Brazil.
In addition, this research has been supported by SINAPAD/Brazil.

\end{document}